# Interpreting the Dimensions of Speaker Embedding Space

*Mark Huckvale*

Speech, Hearing and Phonetic Sciences, University College London, U.K.

`m.huckvale@ucl.ac.uk`

## Abstract

Speaker embeddings are widely used in speaker verification systems and other applications where it is useful to characterise the voice of a speaker with a fixed-length vector. These embeddings tend to be treated as "black box" encodings, and how they relate to conventional acoustic and phonetic dimensions of voices has not been widely studied. In this paper we investigate how state-of-the-art speaker embedding systems represent the acoustic characteristics of speakers as described by conventional acoustic descriptors, age, and gender. Using a large corpus of 10,000 speakers and three embedding systems we show that a small set of 9 acoustic parameters chosen to be "interpretable" predict embeddings about the same as 7 principal components, corresponding to over 50% of variance in the data. We show that some principal dimensions operate differently for male and female speakers, suggesting there is implicit gender recognition within the embedding systems. However we show that speaker age is not well captured by embeddings, suggesting opportunities exist for improvements in their calculation.

## 1. Introduction

Speaker embeddings are high dimensional vectors used to characterise the voice of a talker. They have application within speaker verification systems because of their property that similar voices will give rise to similar vectors, while dissimilar voices will give rise to dissimilar vectors [1]. Speaker embeddings are typically calculated by deep neural network systems trained with multiple recordings of thousands of speakers. The network estimates a probability distribution over known speakers, and the activation of a penultimate bottleneck layer can be used as the speaker embedding. Although widely exploited the high-dimensional space occupied by speaker embeddings has not been widely studied, for example it is not clear what properties of the speech signal are represented in the embeddings. Neither is it clear how the principal dimensions of this space relate to more conventional acoustic descriptors of voices, or how they relate to human perception of speaker similarity. It is possible that a better understanding of how speaker embeddings capture speaker characteristics could lead to improved performance in speaker verification.

Another application for speaker embeddings is within voice conversion, where a recording of one speaker is transformed into the voice of another. Typically such systems use an audio recording of the target speaker to provide the information necessary to create a copy of the source speaker's speech in the target speaker's voice [2]. But recent work in voice conversion has now started to use a speaker embedding of the target voice instead of an audio recording [3]. The input speech is first converted into a phonetic representation (such as a phone posteriorgram) and the synthesis section of the system then takes the phonetic representation and the target speaker embedding to generate audio. The use of a speaker embedding to control output generation assumes that it captures the characteristics of the speech necessary to emulate the target speaker. These characteristics are not necessarily the same as those suited to identifying the speaker, since success at voice conversion will ultimately be based on human listener perception of voice similarity rather than verification. A better understanding of the nature of speaker embedding space could be useful to improve the effectiveness of embeddings for voice conversion. For example, it is likely that embeddings are not good representations of long-term prosody, or how changes in voice quality vary with sentence position since they are based on short segments of speech audio.

There are also novel applications for any-to-any voice conversion when no recordings are available of the target voice, and the user needs to specify the characteristics of the voice using text descriptions or dialog controls. Specific examples include [4]. Some ideas have been put forward in which a latent space of speakers could be created using speaker embeddings, and users could select from a range of speakers organized into a two-dimensional map [5, 6]. Such an application would also benefit from a better understanding of the space of talker variation captured in speaker embeddings, perhaps leading to definitive set of speaker descriptors that could be used to specify a target voice.

This paper will present an analysis of speaker embedding space that starts to unravel how conventional descriptors of voices are related to the principal dimensions of speaker embeddings. By training embeddings and evaluating embeddings using representative samples of speakers, we can start to explore how embeddings vary with age and gender. By computing a number of standard acoustic parameters of the voices, we can see how well embeddings can be predicted from known acoustic properties, and whether the principal dimensions of speaker embedding space have simple acoustic explanations.

## 2. Data and Methods

### 2.1. Datasets

For *training* the speaker embedding systems, we used samples from the following corpora: (i) a set of 921 speakers from **LibriSpeech** - a collection of audiobook recordings [7]; (ii) a set of 1,000 speakers balanced by gender from **VoxCeleb2** - a collection of extracts from YouTube videos [8]; (iii) a set of 1,043 speakers balanced for age and gender from **Globe**, a curated subset of the CommonVoice corpus of speakers donating their voice for science [9, 10].

For *testing* the trained speaker embedding systems we used a representative sample of the Globe corpus containing 4,918 male and 4,918 female speakers as labelled in the corpus metadata. To ensure sufficient audio data for each speaker to get stable mean values of acoustic parameters, we chunked smaller recordings into segments of at least 30s. This gave a total of 33,184 audio files for the 9,836 speakers.

### 2.2. Speaker Embedding Model

We use the Deep-Speaker model to train the speaker embedding systems [11]. This model uses a residual CNN analysis network followed by a pooling layer and a penultimate dense layer to create embeddings. We explored the use of embeddings of size 256 and 512; the results for size 256 are presented here, although size 512 gave similar results. For this study, we increased the size of the input samples to 480 frames of a 64 channel mel-scaled filterbank, representing 4.8s of speech sampled at 16000 samples/sec.

The Deep-Speaker model was trained in two stages as described in the supplied training scripts. Stage 1 trained a softmax model with only a set of positive speaker examples. Stage 2 used triplet training [12] to fine tune the softmax outputs by including negative samples chosen from speakers not included in the training data. The training of an embedding system took approximately 2 days on an Nvidia RTX A5000 GPU. The embeddings computed for the test data were calculated for individual recordings and then the mean was calculated to represent the embedding for each test speaker.

### 2.3. Acoustic Analysis

The testing data samples were analysed using a conventional set of acoustic parameters chosen to have a somewhat transparent interpretation in terms of acoustic-phonetic properties, see Table 1.

Table 1 - Acoustic parameters and interpretation

| Interpretable Voice Property | Physical Parameter | Units |
|---|---|---|
| Pitch height | FXMEDIAN | st |
| Pitch range | FXIQR | st |
| Irregularity | PPQ | % |
| Breathiness | GNE | 0-1 |
| Brightness | SLOPE | dB/kHz |
| Size | VTLEN | cm |
| Loudness | LEVEL | dB |
| Intelligibility | STOI | 0-1 |
| Signal Quality | PESQ | 1-5 |

These properties were computed as follows:
- FXMEDIAN, the median fundamental frequency was calculated from pitch period epochs computed using REAPER [13].
- FXIQR, the fundamental frequency inter-quartile range was also computed from the pitch periods.
- PPQ, the period perturbation quotient was computed from the pitch periods by comparing each pitch period duration to the mean of its neighbours [14]. A large value shows greater irregularity or creakiness in the voice.
- GNE, the glottal noise energy was computed in voiced regions by correlating energy across different frequency bands [15]. A low correlation shows greater noise or breathiness in the voice.
- SLOPE, the spectral slope in voiced regions, was calculated by fitting a line to the spectrum represented as energy in 1000Hz wide filters centred at 1500, 2000, 2500, 3000 and 3500Hz.
- VTLEN, the estimated vocal tract length, was calculated from formant frequencies collected in voiced, vocalic regions of the signal. In places where four formant frequencies could be reliably estimated, these were used to calculate the length of a half-open tube having the best fit to these resonances [16]. The mean length was represented in cm.
- LEVEL, the average level of the signal, was calculated from the RMS value of the recorded signal. Note that since the recordings were uncalibrated, this does not represented dBSPL, but only the arbitrary recording level and speaking level of the recording.
- STOI, a measure of speech intelligibility, was calculated using a non-intrusive estimate from the SQUIM toolbox [17]
- PESQ, a measure of signal quality, was also calculated using a non-intrusive estimate from SQUIM.

These parameters were calculated for all recordings in the test data, and then means taken to represent the average acoustic properties of each speaker. The parameters were transformed and z-scored to ensure each had a relatively normal distribution.

### 2.4. PCA + Greedy Feature Models

Principal components analysis (PCA) was performed using the R statistics system. For PCA of the speaker embeddings, no additional scaling was performed.

Linear model building to predict the principal embedding dimensions from the acoustic parameters was performed using a generalised linear model in R. A greedy feature selection methods was used to find the best parameters for each dimension: first the single parameter which gave the best fit was chosen, then the other parameters were tested in turn to find the best pair of parameters, and then the third and so on. Each repetition was evaluated on a held-out test set, and model building was terminated when the addition of any further parameter did not improve the performance of the model on the held-out data, or if a Chi-Square test showed that the updated model made no significant improvement in prediction.

## 3. Results

### 3.1. PCA analysis from three different training datasets

Embeddings for the Globe test data of 9836 speakers were calculated using the three versions of the embedding system trained on LibriSpeech, VoxCeleb2 and Globe. The positions of the test speakers in the first two principal dimensions for each system are plotted in Fig.1. Note that similar plots are obtained for the systems, suggesting that there is some common structure to the embedding space independent of the training speakers, at least for the first two dimensions. We will focus on the system trained on the representative Globe subset for the rest of the paper

We can look at how well the PCA represents the structure of the embedding space by computing the amount of variance in the embeddings captured by an increasing number of components. We can also ask how well the embeddings computed from a constrained number of components match the original embeddings – this can be calculated from the mean cosine distance between the actual embeddings and those reconstructed from the principal components. Table 2 shows

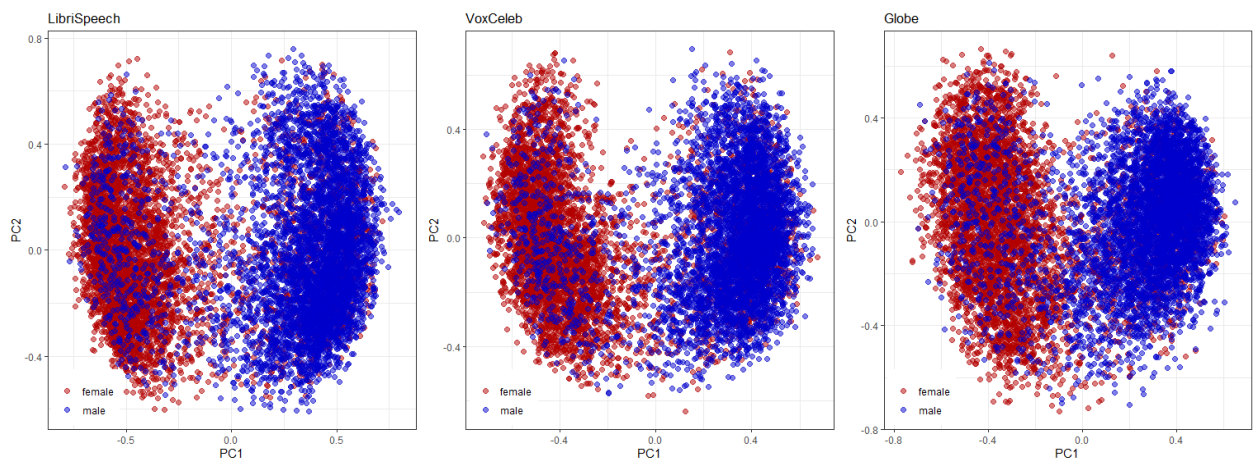

Figure 1 - First two principal component dimensions for the same test data from a speaker embedding system trained with 3 different datasets. Note that the sign of each dimension has been chosen to be consistent. Colours reflect gender labels given in Globe dataset.

how % variance explained and mean cosine distance varies with number of components used.

Table 2 - Quality of PCA fit with # components included

| # Components | % Variance explained | Mean cosine distance |
|---|---|---|
| 1 | 15.7 | 0.657 |
| 2 | 23.4 | 0.570 |
| 3 | 29.8 | 0.510 |
| 4 | 35.8 | 0.460 |
| 5 | 41.3 | 0.415 |
| 6 | 46.2 | 0.377 |
| 7 | 50.4 | 0.349 |
| 8 | 54.2 | 0.323 |
| 9 | 57.8 | 0.300 |
| 10 | 60.9 | 0.281 |

Ten principal components capture 60% of the variance in the embeddings, while 30 components are required to capture 90% of the variance. In terms of cosine distance, we first note that the mean interspeaker distance in the test set is 0.85, while 99% of all interspeaker distances are greater than 0.34. A model with 8 principal components can estimate embeddings within a cosine distance of 0.323 on average, which is smaller than most inter-speaker distances. While that would likely be insufficient for a speaker verification system, it shows that a system with 8 components is capturing some useful properties of the embedding space.

### 3.2. Acoustic analysis of PCs

To determine how well the embedding space can be predicted from conventional acoustic parameters, we train an MLP regression model to predict embeddings from the normalised acoustic parameters. The MLP has 9 inputs, hidden units of 32 and 128, and 256 outputs and was trained using the `sklearn` python package. Performance is estimated on the Globe test data using 5-fold cross validation. The set of 9 acoustic parameters predicts the embeddings with a mean cosine distance of 0.354, equivalent to about 7 principal components. This suggests that these parameters are may be useful for the characterisation of a target speaker in voice conversion.

### 3.3. Modelling of gender in embedding space

The PCA analyses in Fig.1 show clear clustering by gender, but also a lot of speakers which appear to be male speakers that sound like female speakers or vice versa. In fact listening to audio samples of the mismatched speakers, it is clear that the problem arises from faulty labels in the Globe corpus. The mismatched speakers "sound" like their position in the plot, not like the gender labels. We can apply a simple clustering in the first two principal dimensions using a 2D Gaussian Mixture Model to approximately relabel the genders. Only 83% of the Globe labels match the cluster labels.

Having relabelled the speakers' gender, we can now ask how many of the principal dimensions are sensitive to gender through being clearly bimodal. In fact only the first principal component is clearly bimodal, and other dimensions do not show a marked differentiation according to gender.

### 3.4. Acoustic interpretations of principal components

To understand how the principal dimensions relate to the set of conventional acoustic parameters, we build a greedy linear model to predict each component. Table 3 shows the parameters chosen as the best for explaining each dimension. The RMSE and Corr columns show the RMS error of prediction and the correlation of the prediction to the actual component value. Note that only the lower dimensions are predicted well, and that all acoustic parameters are used somewhere in the prediction of the first 8 components. In general the mappings between acoustic parameters and principal components is complex except for PC1.

The analysis in Table 3 assumes that the best acoustic model is the same for both male and female speakers. If we repeat the modelling, but build separate models for male and female speakers (as found by clustering) then we can see significant improvements in the quality of prediction, particularly for certain dimensions. Table 4 show the RMSE and Correlation figures for models that are calculated with and without regard to gender. Particular improvements are seen for dimensions 2 and 8. The gendered modelling for these dimensions also have a simpler explanation in terms of the greedily-chosen acoustic parameters.

Table 3 - Best linear model of acoustic parameters for PCA dimensions

| PC# | Acoustic parameters | RMSE | Corr |
|---|---|---|---|
| 1 | FXMEDIAN, VTLEN, PESQ | 0.128 | 0.936 |
| 2 | PESQ, GNE, PPQ, STOI, FXMEDIAN, FXIQR, VTLEN, SLOPE | 0.236 | 0.336 |
| 3 | SLOPE, PPQ, PESQ, FXMEDIAN, VTLEN, GNE, FXIQR | 0.201 | 0.497 |
| 4 | GNE, LEVEL, PESQ, SLOPE, VTLEN, FXMEDIAN, PPQ | 0.200 | 0.430 |
| 5 | LEVEL, GNE, FXMEDIAN, VTLEN, PESQ, STOI | 0.205 | 0.281 |
| 6 | FXMEDIAN, VTLEN, GNEMIN | 0.194 | 0.189 |
| 7 | SLOPE, FXMEDIAN, VTLEN, LEVEL | 0.175 | 0.269 |
| 8 | PPQ, LEVEL, GNE, PESQ, VTLEN | 0.174 | 0.193 |

Table 4 – Acoustic model fit when split by gender

| PC# | Non-Gendered | | Gendered | |
|---|---|---|---|---|
| | RMSE | Corr | RMSE | Corr |
| 1 | 0.128 | 0.936 | 0.110 | 0.953 |
| 2 | 0.236 | 0.336 | **0.211** | **0.553** |
| 3 | 0.201 | 0.497 | 0.175 | 0.664 |
| 4 | 0.200 | 0.430 | 0.195 | 0.468 |
| 5 | 0.205 | 0.281 | 0.178 | 0.470 |
| 6 | 0.194 | 0.189 | 0.190 | 0.369 |
| 7 | 0.175 | 0.269 | 0.167 | 0.413 |
| 8 | 0.174 | 0.193 | **0.149** | **0.547** |

Table 5 - Examples of gendered dimensions

| PC# | Gender | Effect of FXMEDIAN | Effect of VTLEN |
|---|---|---|---|
| 1 | Both | ↓ | ↑ |
| 2 | Female | ↑ | ↓ |
| 2 | Male | ↓ | ↑ |
| 8 | Female | ↑ | ↑ |
| 8 | Male | ↓ | ↓ |

A closer look at the best gendered models for these dimensions show that while similar parameters are found in the model for both male and female speakers, the sign of the contribution of the parameter to the prediction can change. Table 5 give some examples for how the numerical effect of pitch and vocal tract length on the component changes sign between male and female models. Overall, this shows that gender plays a significant role in how the acoustic properties of the recording relate to a speaker embedding.

### 3.5. Modelling of age in embedding space

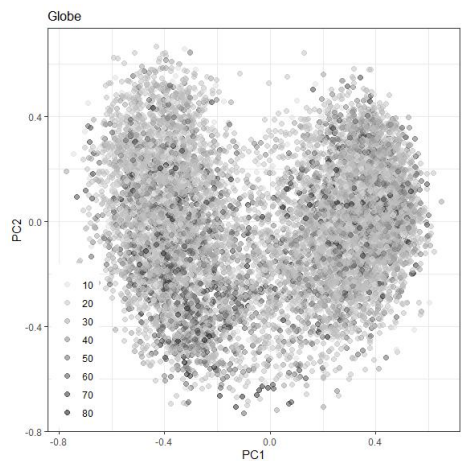

Figure 2 - Representation of Age in embedding space

While there are clear indications of gender sensitivity in the embedding space, the representation of age is much less clear. Figure 2 shows the first two dimensions of the PCA coloured by decade. None of the first 10 principal components correlate better than 0.15 with age. Prediction of age from the embeddings only shows a correlation of 0.29 on held-out data.

## 4. Conclusions

In this study we have investigated the properties of the dimensions of the space of speaker embeddings. We have shown that differently trained embeddings have similar principal dimensions after PCA (at least for first few dimensions), showing that there is some underlying structure of speaker embedding space that has the potential for interpretation. We have also looked at some pre-trained embeddings and obtained similar results for PC1&2 to those shown here.

The 9 interpretable dimensions tested here can predict embeddings about as accurately as 7 principal components, so may be useful for target speaker definition in voice conversion. There are many opportunities for improving these dimensions, either by increasing their explanatory power or by improving their interpretability. In particular the use of PPQ and GNE for creakiness and breathiness seems weak, and better parameters could be devised, which are both computable from the audio signal and highly correlated with listener perceptions of voice quality.

Only PC1 was found to be significantly bimodal – and unsurprisingly highly correlated with acoustic parameters such as FXMEDIAN and VTLEN which are themselves bimodal. Later PCs look unimodal, which suggests they express a single acoustic space for male and female speakers. The fact that some PCs work differently for male and female suggests that embeddings implicitly encode gender.

Age was not well indicated within the embedding space and age was not found to be a useful predictor of its principal dimensions. This is surprising in that age ought to be a fixed characteristic of a speaker useful for verification.

We also found that the gender labels in Globe are likely to be faulty.

A demonstration of voice conversion using the results in this paper can be found in [18].